\documentclass[aps, preprint]{revtex4}

\usepackage{amsfonts}
\usepackage{amsmath}
\usepackage{amssymb}
\usepackage{graphicx}
\usepackage{epsfig}
\providecommand{\U}[1]{\protect\rule{.1in}{.1in}}
\begin{document}
\title{MODELLING OF NONLOCAL EFFECTS IN ELECTROMECHANICAL NANO-SWITCHES}
\author{M.M. Toropova}
\affiliation{Department of Mechanical and Industrial Engineering, University of Toronto,
Toronto, Canada}
\email{marina.toropova@utoronto.ca}
\date{\today}

\begin{abstract}
Dielectric nano-swithes made of the materials that exhibit piezoelectric
and/or flexoelectric properties with significant electro-mechanical coupling
are considered. In this case, a nonuniform strain field may locally break
inversion symmetry and induce polarization even in nonpiezoelectrics. At
reducing dimensions to the nanoscale, the flexoelectric effect demonstrates
the nonlocality of the dielectric materials and plays more significant role
than piezoelectric effect. The flexoelectric effect is included into
consideration via additional term coupling strain gradient and polarization in
the electric enthalpy density. The equations of motion of the improved
Euler-Bernoulli and Timoshenko beam models, and 2-D plate theory have been obtained.

\end{abstract}
\maketitle

%%%%%%%%%%%%%%%%%%%%%%%%%%%%%%%%%%%%%%%%%%%%%%%%%%%%%%%%%%%%%%%%%%%%%%

%%%%%%%%%%%%%%%%%%%%%%%%%%%%%%%%%%%%%%%%%%%%%%%%%%%%%%%%%%%%%%%%%%%%%%
\section{Introduction}

Electromechanical nano-switches as integral parts of nanoelectromechanical systems (NEMS) find numerous technological applications as, e.g., mass memory storage, high-frequency electrical switches, and mass or force sensors. Such high technological applications demand combined efforts of engineering and science through modelling and simulations. Hence, to design state-of-the-art nanotechnological devices with predetermined characteristics, engineers need in addition to experimental techniques not only new formulas and computational methods but also improved electromechanical models. The new models must account for the nonlocal properties of the materials and new physical phenomena.

The nonlocality appears due to the noticeable role of interatomic forces in nano-objects. In nonlocal theories, the constitutive equations take into account microstructure of real materials and microscopic interaction length between, e.g., molecules in a lattice. From the physical point of view, the nonlocality exhibits through special size-dependent effects such as, for example, flexoelectricity - induced polarization due to the strain gradient. It is well known that conventional continuum theories are size-independent and therefore cannot be applied automatically for the analysis of NEMS devices. The most suitable tools for their analysis are atomic and molecular models, but they are restricted by their computational capacity. One of the ways to resolve this contradiction consists in the employment of improved, size-dependent classical theories [1].

In the present work, we consider both piezoelectric and flexoelectric cases. In the latter case, a nonuniform strain field may locally break inversion symmetry and induce polarization. To take into account this effect, similar to [1], we assume that the electric enthalpy density depends not only on strain, electric field, and polarization but also on  strain gradient. Then, we replace the polarization  by its linear representations through the electric field and strain. Using Hamilton's principle, we obtain the equations of motion for  Euler-Bernoulli and Timoshenko beams as well as for 2-D plate. The higher order terms in the strain from the electric enthalpy density give  additional contribution to the bending rigidity of the beam compared to classical solution.  The analysis shows that taking into account the strain gradient increases the elastic characteristics of the nano-switch  considered in this work. From the obtained formulas  it is seen that at reducing dimensions to the nanoscale the flexoelectric effect plays more significant role than piezoelectric effect.  In the dynamic case, the Euler-Bernoulli beam model provides overestimated frequencies. Two-dimensional model allows us to evaluate the electric potential accumulated in the nano-switch due to induced nonuniform strain.

%%%%%%%%%%%%%%%%%%%%%%%%%%%%%%%%%%%%%%%%%%%%%%%%%%%%%%%%%%%%%%%%%%%%%%

%%%%%%%%%%%%%%%%%%%%%%%%%%%%%%%%%%%%%%%%%%%%%%%%%%%%%%%%%%%%%%%%%%%%%%
\section{Governing Equations}
Electromechanical nano-switch may be simulated as a dielectric cantilever nanobeam (Fig. 1). The material of the beam may have tetragonal or cubic symmetry or in another words it may be piezo- or nonpiezoelectric.

First, let us consider flexoelectric properties of the beam. In \cite{1}-\cite{a3} the fourth-order flexoelectric tensor is introduced in two different ways: as $\mu_{ijkl}u_{j,kl}$
\begin{eqnarray}
P_{l}=\varepsilon_{0}\eta_{lm}E_{m}+e_{ljk}S_{jk}+\mu_{ijkl}u_{i,jk}
			\label{eq:cc1}
\end{eqnarray}
or as $f_{ijkl}$
 in the additional term in the thermodynamic potential \cite{a3}:
\begin{eqnarray} 
f_{ijkl}P_{i}u_{j,kl}.		\label{eq:ccc1}
\end{eqnarray}

Here, $S_{ij}$, $e_{ijk}$, $\eta_{lm}$,    are the components of the strain, dielectric, and relative permittivity  tensors,  respectively, $P_{i}$, $u_{i}$,  $E_{m}$, are the components of polarization, displacement, and electric field vectors, respectively, $\varepsilon_o$ is the permittivity of vacuum.

In \cite{a4} there is a structure of the flexoelectric coefficients for a cubic material (in matrix notation):
\begin{eqnarray}
\left(
\begin{array}{cccccc}
\mu_{11} & \mu_{12} & \mu_{12} & 0 & 0 & 0\nonumber  \\ 
\mu_{12} & \mu_{11} & \mu_{12} & 0 & 0 & 0 \nonumber  \\ 
\mu_{12} & \mu_{12} & \mu_{11} & 0 & 0 & 0 \nonumber  \\ 
 0 & 0 & 0 & \mu_{44} & 0 & 0\nonumber  \\  
 0 & 0 & 0 & 0 & \mu_{44} & 0 \nonumber \\
 0 & 0 & 0 & 0 & 0 &\mu_{44}  \nonumber
\end{array}
\right).
\end{eqnarray}

\begin{figure}
\begin{center}
\epsfig{file=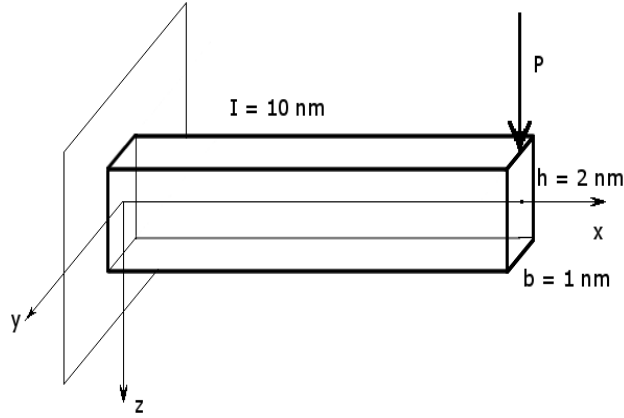, height=6.8cm, width=8.9cm}
\caption{Cantilever beam(initial configuration).}
\label{f1} 
\end{center}
\end{figure}
  
According to \cite{1} and \cite{2}, the electric enthalpy may be presented in the form
\begin{eqnarray}
&&H(S_{ij}, P_{i}, P_{i,k}, u_{i,jk})=W(S_{ij}, P_i, P_{i,k}, u_{i,jk})-\frac{1}{2}\varepsilon_{0}\varphi_{,i}\varphi_{,i}+
\varphi_{,i}P_{i}, \label{eq:c1}
\end{eqnarray}
\begin{eqnarray}
&& W(S_{ij}, P_i, P_{i,k}, u_{i,jk})=\frac{1}{2}a_{ij}P_i P_j +\frac{1}{2}b_{ijkl}P_{j,i}P_{l,k}+\frac{1}{2}c_{ijkl}S_{ij}S_{kl}\nonumber	 \\ &&+d_{ijk}S_{jk}P_{i}+  	
 e_{ijkl}P_{j,i}S_{kl}
+f_{ijkl}P_{i}u_{j,kl}+g_{ijk}P_{i}P_{k,j},	\label{eq:c2}	
\end{eqnarray}
where $W$ is an energy density of deformation and polarization,  $a_{ij}$, $b_{ijkl}$, $c_{ijkl}$, $d_{ijk}$, $e_{ijkl}$, $f_{ijkl}$, $g_{ijk}$ are the components of reciprocal dielectric susceptibility, polarization gradient-polarization gradient coupling tensor, elastic tensor, piezoelectric tensor, polarization gradient-strain coupling tensor, flexoelectric tensor, and polarization-polarization gradient coupling tensor, respectively, $\varphi$ is the potential of electric field.

Since the components of the stress tensor and the electric field vector $E_{i}=-\varphi^{'}_{i}$ are expressed as 
\begin{eqnarray}
T_{ij}=\frac{\partial W}{\partial S_{ij}}, \quad 
E_{i}=-\frac{\partial W}{\partial P_{i}},	\nonumber	
\end{eqnarray}
we have
\begin{eqnarray}
T_{ij}=c_{ijkl}S_{kl}+d_{ijk}P_{k}+e_{ijkl}P_{k,l}  \nonumber\\
-E_{i}=a_{ij}P_{j}+d_{ijk}S_{jk}+f_{ijkl}u_{i,jk}  \label{eq:c3}
\end{eqnarray}
Compare (\ref{eq:cc1}), (\ref{eq:ccc1}), and the second equation of (\ref{eq:c3}), we conclude that flexoelectric tensor  $f_{ijkl}$ has 
21 nonzero components: 
\begin{eqnarray}
f_{1111}=-\frac{\mu_{11}}{a_{11}}, \quad 
f_{2211}=f_{3311}=-\frac{\mu_{12}}{a_{11}}  \nonumber\\
f_{1221}=f_{2121}=f_{1331}=f_{3131}=-\frac{\mu_{44}}{a_{11}} \nonumber\\
f_{2222}=-\frac{\mu_{11}}{a_{22}}, \quad 
f_{1122}=f_{3322}=-\frac{\mu_{12}}{a_{22}}  \nonumber\\		
f_{1212}=f_{2112}=f_{2332}=f_{3232}=-\frac{\mu_{44}}{a_{22}} \nonumber \\
f_{3333}=-\frac{\mu_{11}}{a_{33}}, \quad 
f_{1133}=f_{2233}=-\frac{\mu_{12}}{a_{33}} \nonumber\\	
f_{1313}=f_{3113}=f_{2323}=f_{3223}=-\frac{\mu_{44}}{a_{33}} \nonumber
\end{eqnarray}
In this work, we do not consider the terms with $b_{ijkl}$, $e_{ijkl}$, and $g_{ijk}$ coefficients although it might be done without difficulties.

Using the equations (\ref{eq:c1})-(\ref{eq:c3}) we derive
 the equations of motion for a dielectric nanoswitch.

Here we consider a cantilever switch which may be presented as a Euler-Bernoulli or Timoshenko beam or 2-D narrow plate (Fig.~\ref{f1}). The load applied to the free edge of the cantilever simulates the case of nonuniform strain.

%%%%%%%%%%%%%%%%%%%%%%%%%%%%%%%%%%%%%%%%%%%%%%%%%%%%%%%%%%%%%%%%%%%%%%
\section{Euler-Bernoulli beam model}
Within the framework of the Euler-Bernoulli model, the displacement can be written as
\begin{eqnarray}
&& u_{1}(x,y,z,t)\equiv u\approx -zw^{'}(x,t), \quad	
u_{2}(x,y,z,t)\approx 0, \quad \nonumber	 \\
&& u_{3}(x,y,z,t)\approx w(x,t),			\label{eq:c0}
\end{eqnarray}
Hence, based on the equations (\ref{eq:c0}), the strains are equal to
\begin{eqnarray}
S_{11}\equiv\varepsilon_x=-zw^{''}, \quad 
S_{22}\equiv\varepsilon_y=0, \quad
S_{33}\equiv\varepsilon_z=0, \quad \nonumber\\	
S_{12}=S_{21}\equiv\gamma_{xy}=S_{13}=S_{31}\equiv\gamma_{xz}=S_{23}=S_{32}\equiv\gamma_{yz}=0.		\label{eq:c5}
\end{eqnarray}
We assume that the electric field acts only in $z-$direction. Then,  from (\ref{eq:c3}) we can express $E_z$ as
\begin{eqnarray}
&&-E_{3}\equiv -E_z =a_{33}P_3+d_{31}\varepsilon_x +f_{1133}u_{,xz}+f_{1313}u_{,zx}+f_{3113}w_{,xx}= \nonumber \\
&&a_{33}P_z-(d_{31}z -f_{1133})w^{''}\nonumber
\end{eqnarray}
or
\begin{eqnarray}
P_{3}\equiv P_z=\frac{1}{a_{33}}(-E_z +(f_{1133}+d_{31}z) w^{''}),\label{eq:c6}
\end{eqnarray}
where the low letter indexes after a comma mean the derivation with respect to corresponding coordinates.

After substituting the equations (\ref{eq:c5}), (\ref{eq:c6}), into 
(\ref{eq:c1}) and (\ref{eq:c2}) and ignoring the polarization gradient, we have
\begin{eqnarray}
&& H=\frac{1}{2}c_{11}z^2 (w^{''})^2+\frac{1}{2a_{33}}(\varphi_{z}^{'} +(f_{1133}+d_{31}z)w^{''})^2- \nonumber \\ 	
&& d_{31}w^{''}\frac{z}{a_{33}}(\varphi_{z}^{'} +(f_{1133}+d_{31}z)w^{''})
-f_{1133}\frac{w^{''}}{a_{33}}(\varphi_{z}^{'} +(f_{1133}+d_{31}z)w^{''}) \nonumber \\ 	
&&-\frac{\varepsilon_0}{2}E_{z}^{2}-\frac{\varphi_{z}^{'}}{a_{33}}(\varphi_{z}^{'} +(f_{1133}+d_{31}z)w^{''})			\label{eq:c7}
\end{eqnarray}
The equation of motion of the beam are derived via Hamilton's principle \cite{a5}:
\begin{eqnarray}
\delta\int_{t_1}^{t_2} (T_{k}-H+W_{d})dt=0				\label{eq:c8}
\end{eqnarray}
where $\delta(\cdot)$ denotes the first variation, $T_{k}$ is the kinetic energy,  and $W_{d}$ is the work done by the external forces and moments.

The kinetic energy of the beam by using the equation (\ref{eq:c0}) can be written as 
\begin{eqnarray}
&&T_{k}=\frac{1}{2}\int_{V_b} \varrho\lbrace\dot{u}\rbrace^{T}\lbrace\dot{u}\rbrace dV_b=
\frac{1}{2}\int_{V_b} \varrho[\dot{w}^2+z^{2}(\dot{w}^{'})^2]dV_b=
\nonumber \\
&&\frac{1}{2}\varrho
\int_{0}^{l}[(\dot{w})^2 A+I(\dot{w}^{'})^2]dx,	\label{eq:c9}		
\end{eqnarray}
where
$I=\int_A z^2 dA$, $A$ is the area of the beam's cross section, $V_{b}$ is the volume of the beam,  $\varrho$ is the mass density of the nano-switch's material, and the length of the beam is equal to $l$.

If the beam is under the transverse force $q$, the work done is equal to
\begin{eqnarray}
W_{d}=\int_{0}^{l}qwdx.								\label{eq:000}
\end{eqnarray}

Performing variation in (\ref{eq:c8}) and using (\ref{eq:c7}), (\ref{eq:c9}), and (\ref{eq:000}), we obtain the equations describing the electromechanical behavior of the beam
\begin{eqnarray}
&&w^{IV}(c_{11}I -\frac{d_{31}^2 I}{a_{33}}-\frac{f_{1133}^2}{a_{33}}A)-\nonumber \\
&&\frac{1}{a_{33}}\frac{\partial^2}{\partial x^2}(f_{1133}Q^{el}_{z}+d_{31}M^{el}_{z})+
\varrho A \ddot{w}+\varrho I\frac{\partial^4 w}{\partial x^2 \partial t^2}=q,							\label{eq:c10}
\end{eqnarray}
\begin{eqnarray}
&&(\frac{3}{a_{33}}-\varepsilon_{0})\varphi_{zz}^{''}+\frac{d_{31}}{a_{33}}w^{''}=0				\label{eq:cc10}
\end{eqnarray}
where $Q^{el}_{z}$ and $M_{z}^{el}$ are the electric transverse shear force and the electric bending moment, respectively:
\begin{eqnarray}
Q^{el}_{z}=\int_{A}E_{z}(x,z)dA \nonumber \\
M^{el}_{z}=\int_{A}zE_{z}(x,z)dA \nonumber
\end{eqnarray}
with the associated boundary conditions \\
at $x=0$
\begin{eqnarray}
w=w^{'}=0,  					\label{eq:001}
\end{eqnarray}
at $x=l$	
\begin{eqnarray}
&&(c_{11}I -\frac{d_{31}^2 I}{a_{33}}-\frac{f_{1133}^2}{a_{33}}A)w^{''}-\frac{d_{31}}{a_{33}}M_{z}^{el}-\frac{f_{1133}}{a_{33}}Q_{z}^{el}=0, \nonumber \\
&&(c_{11}I -\frac{d_{31}^2 I}{a_{33}}-\frac{f_{1133}^2}{a_{33}}A)w^{'''}-\frac{d_{31}} {a_{33}}\frac{\partial}{\partial x}M_{z}^{el}-     \label{eq:002}  \\
&&\frac{f_{1133}}{a_{33}}\frac{\partial}{\partial x}Q_{z}^{el}=P_{*}, 		 \nonumber	
\end{eqnarray}
and at $z=\pm h/2$
\begin{eqnarray}
\varphi^{'}_{z}=0,  					\label{eq:00002}
\end{eqnarray}
where $P_{*}$ is a force applied to the right edge of the cantilever.

The equations (\ref{eq:c10}),  (\ref{eq:cc10}) and the boundary conditions (\ref{eq:001}) -  (\ref{eq:00002}) couple the mechanical displacement $w$ and the electric potential $\varphi$.

From the equation (\ref{eq:c10}) it is seen that piezoelectric and flexoelectric effects increase the bending rigidity of the beam ($a_{33}<0$) and  for the beams with large  cross-section area and consequently large its moment of inertia, the term $Af_{1133}^2/ a_{33}$ plays insignificant role. However, for small cross-sectional dimensions, when 
$$\frac{f_{1133}^2}{a_{33}}>(c_{11} -\frac{d_{31}^2}{a_{33}})\frac{h^2}{12},$$
where $h$ is the thickness of the beam, the flexoelectric effect becomes noticeable.
 
 The normalized Young modulus is
$$Y=1-\frac{d_{31}^2}{c_{11}a_{33}}-\frac{12 f_{1133}^2}{c_{11}a_{33}h^2},$$
It is seen that at $h\rightarrow 0$ the flexoelectric term plays the dominant role in the bending rigidity.
 
The classical formula for the maximum delection of a cantilever beam $w_{max}=P_{*}l^3/3Y$ may be used to find the coefficient $f_{1133}$ from experimental results:
\begin{eqnarray}
f_{1133}=\sqrt{\frac{1}{a_{33}}[c_{11}\frac{h^2}{12}-\frac{P_{*}l^3}{3bhw_{max}.
}]} \nonumber
\end{eqnarray}

It is obvious that the dispersive relation is
\begin{eqnarray}
\omega^2 \varrho( A -I k^2)-k^4(c_{11}I -\frac{d_{31}I}{a_{33}}-\frac{f_{1133}^2}{a_{33}}A)=0   \label{eq:0002}
\end{eqnarray}
where $\omega$ is a frequency and $k$ is a wavenumber.
% !!!!!!!!!!!!!!!!!!!!!!!!!!!!!!!!!!!!!!!!!!!!!!!!!!!!!!!!!!!!!!!!+
\section{Timoshenko beam model}

In this model we assume that the displacements are presented as
\begin{eqnarray}
&&u_{1}(x,y,z,t)\approx -z\psi(x,t), \quad	
u_{2}(x,y,z,t)\approx 0, \quad  \nonumber	 \\
&&u_{3}(x,y,z,t)\approx w(x,t),			\label{eq:b1}
\end{eqnarray}
where $w$ is the transverse displacement of the points of the centroidal axis($y=z=0$), and $\psi$ is the rotation of the beam cross-section about the positive $y-$axis.

From (\ref{eq:b1}), the strains are equal to
\begin{eqnarray}
&&\varepsilon_x=-z\psi^{'}, \quad 
\varepsilon_y=0, \quad
\varepsilon_z=0, \quad \nonumber	 \\
&&\gamma_{xy}=\gamma_{yz}=0, \quad
\gamma_{xz}=w^{'}-\psi.	
										\label{eq:b2}
\end{eqnarray}

Since the narrow beam deflects in $x-z$ plane, we consider  only two components of the electric field $E_x$ and  $E_z$. From (\ref{eq:c3}) they can be expressed  as
\begin{eqnarray}
&&-E_x =a_{11}P_x +d_{15}\gamma_{xz}+f_{1111}u_{,xx} \nonumber \\
&&-E_z =a_{33}P_z +d_{31}\varepsilon_{x}+f_{1133}u_{,xz}+f_{1313}u_{,zx}+f_{3113}w_{,xx},
												\nonumber
\end{eqnarray}
or
\begin{eqnarray}
&&-E_x =a_{11}P_x +d_{15}(w^{'}-\psi)-f_{1111}z\psi^{''} \nonumber \\
&&-E_z =a_{33}P_z -d_{31}z\psi^{'}-f_{1133}\psi^{'}+f_{1313}(w^{''}-\psi^{'}),
											\nonumber	
\end{eqnarray}
and hence
\begin{eqnarray}
&&P_x=\frac{1}{a_{11}}(-E_x -d_{15}(w^{'}-\psi)+f_{1111}z\psi^{''}), \nonumber \\
&&P_z=\frac{1}{a_{33}}(-E_z +d_{31}z \psi{'}+f_{1133}\psi^{'}-f_{1313}(w^{''}-\psi^{'})),															\label{eq:b4}
\end{eqnarray}

After substituting the equations (\ref{eq:b2}), (\ref{eq:b4}) into 
(\ref{eq:c1}) and (\ref{eq:c2}) and ignoring the polarization gradient, we have
\begin{eqnarray}
&& H=\frac{1}{2}a_{11}P_{x}^2 +\frac{1}{2}a_{33}P_{z}^2+\frac{1}{2} c_{11}\varepsilon_{x}^2+\frac{1}{2}c_{55}\gamma_{xz}^2+d_{15}\gamma_{xz}P_{x}  \nonumber	 \\
&&+d_{31}\varepsilon_{x}P_{z}+f_{1111}P_{1}u_{xx}+
P_{3}(f_{1133}u_{xz}+f_{1313}u_{zx}+f_{3113}w_{xx})-\nonumber	 \\
&&
\frac{1}{2}\varepsilon_0 (\varphi^{'}_{x})^2-\frac{1}{2}\varepsilon_0 (\varphi^{'}_{z})^2+\varphi^{'}_{x}P_{x}+\varphi^{'}_{z}P_{z}= 
\nonumber \\
&&\frac{1}{2a_{11}}(\varphi^{'}_x -d_{15}(w^{'}-\psi)+f_{1111}z\psi^{''})^2 +\nonumber \\
&&\frac{1}{2a_{33}}(\varphi^{'}_z +d_{31}z \psi{'}+f_{1133}\psi^{'}-f_{1313}(w^{''}-\psi^{'}))^2+ \nonumber	 \\
&&\frac{1}{2}c_{11}z^2 (\psi^{'})^2+ \frac{1}{2}c_{55}(w^{'}- \psi)^2+  
d_{15}(w^{'}-\psi)\frac{1}{a_{11}}
(\varphi^{'}_{x}-d_{15}(w^{'}-\psi)+f_{1111}z\psi^{''})-  \nonumber	 \\
&&\frac{d_{31}z\psi^{'}}{a_{33}}
(\varphi^{'}_z +d_{31}z \psi{'}+f_{1133}\psi^{'}-f_{1313}(w^{''}-\psi^{'}))-   \nonumber \\
&&f_{1111}\frac{z\psi^{''}}{a_{33}}(\varphi^{'}_{x}-d_{15}(w^{'}-\psi)+f_{1111}z\varphi^{''}) \nonumber \\
&&f_{1133}\frac{\psi^{'}}{a_{33}}(\varphi^{'}_z +d_{31}z \psi{'}+f_{1133}\psi^{'}-f_{1313}(w^{''}-\psi^{'}))+ \nonumber	 \\
&&f_{1313}\frac{w^{''}-\psi^{'}}{a_{33}}(\varphi^{'}_z +d_{31}z \psi{'}+f_{1133}\psi^{'}-f_{1313}(w^{''}-\psi^{'}))	\nonumber	 \\
&&-\frac{1}{2}\varepsilon_0 (\varphi^{'}_{x})^2-\frac{1}{2}\varepsilon_0 (\varphi^{'}_{z})^2+ 
\frac{\varphi^{'}_{x}}{a_{11}}(\varphi^{'}_x -d_{15}(w^{'}-\psi)+f_{1111}z\psi^{''})+\nonumber	 \\
&&\frac{\varphi^{'}_{z}}{a_{33}}(\varphi^{'}_z +d_{31}z \psi{'}+f_{1133}\psi^{'}-f_{1313}(w^{''}-\psi^{'}))	\nonumber	
\end{eqnarray}
The kinetic energy is equal to
\begin{eqnarray}
&& T_{k}=\frac{1}{2}\int_{V_b} \varrho\lbrace\dot{u}\rbrace^{T}\lbrace\dot{u}\rbrace dV_b=\frac{1}{2}\int_{V_b} \varrho[z\dot{\psi^2}+\dot{w}^2]dV_b=
\frac{1}{2}\varrho\int_{0}^{l}[I(\dot{\psi}^2)+A(\dot{w})^2]dx,  \nonumber
\end{eqnarray}
The work done is described by the equation (\ref{eq:000}).

Applying the Hamilton's principle similar to previous section, we obtain the equations of motion for the Timoshenko beam
\begin{eqnarray}
&& kA(c_{55}-\frac{d_{15}^2}{a_{11}})(w^{''}-\psi^{'})
+\frac{d_{15}}{a_{11}}\frac{\partial}{\partial x}Q_{x}^{el}+\nonumber \\
&&\frac{f_{1313}}{a_{33}}(-\frac{\partial^2}{\partial x^2}Q_z^{el}-kA(f_{1133}\psi^{'''}-f_{1313}(w^{IV}-\psi^{'''})))-q=
\varrho A\ddot{w} \nonumber \\
&& (c_{11}I -\frac{d_{31}^2}{a_{33}}I -\frac{(f_{1133}+f_{1313})^2}{a_{33}}kA)\psi^{''}
+kA(c_{55}-\frac{d_{15}^2}{a_{11}})(w^{'}-\psi)-\nonumber \\
&&\frac{d_{31}}{a_{33}}\frac{\partial}{\partial x}M_{z}^{el}+
\frac{d_{15}}{a_{11}}Q_{x}^{el}+
\frac{f_{1133}+f_{1313}}{a_{33}}(f_{1313}kA w^{'''}-\frac{\partial}{\partial x}Q_{x}^{el})+\nonumber	 \\		
&&+\frac{f_{1111}}{a_{11}}\frac{\partial^2}{\partial x^2}M_{x}^{el}+\frac{f_{1111}^2}{a_{11}}I\psi^{IV}=\varrho I \ddot{\psi}			\nonumber \\ 									
&& \varphi_{xx}(\frac{3}{a_{11}}-\varepsilon_0)+
\varphi_{zz}(\frac{3}{a_{33}}-\varepsilon_0)+(\frac{d_{15}}{a_{11}}+
\frac{d_{31}}{a_{33}})\psi^{'}-
\frac{d_{15}}{a_{11}}w^{''}=0						\nonumber		
\end{eqnarray}
with boundary conditions:\\
at $x=0$	
\begin{eqnarray}
w=w^{'}=\psi=\psi^{'}=\varphi=0,					\nonumber
\end{eqnarray}
and at $x=l$:
\begin{eqnarray}
&& kA(c_{55}-\frac{d_{15}^2}{a_{11}})(w^{'}-\psi)
+\frac{d_{15}}{a_{11}}Q_{x}^{el}+\nonumber \\
&&\frac{f_{1313}}{a_{33}}(-\frac{\partial}{\partial x}Q_z^{el}-kA(f_{1133}\psi^{''}-f_{1313}(w^{'''}-\psi^{''})))=P_{*}
											\nonumber
\end{eqnarray}
\begin{eqnarray}
&& (c_{11}I -\frac{d_{31}^2}{a_{33}}I -\frac{(f_{1133}+f_{1313})^2}{a_{33}}kA)\psi^{'}-
\frac{d_{31}}{a_{33}}M_{z}^{el}+\nonumber	 \\		
&&\frac{f_{1133}+f_{1313}}{a_{33}}(f_{1313}kA w^{''}- Q_{x}^{el})+	
\frac{f_{1111}}{a_{11}}\frac{\partial}{\partial x}M_{x}^{el}+\frac{f_{1111}^2}{a_{11}}I\psi^{'''}=0,   \nonumber 
\end{eqnarray}
\begin{eqnarray}
kA[(f_{1133}+f_{1313})\psi^{'}-f_{1313}w^{''}]+Q_{z}^{el}=0,
													\nonumber
\end{eqnarray}
\begin{eqnarray}
f_{1111}I\psi^{''}+M_{x}^{el}=0,     
											\nonumber
\end{eqnarray}  
\begin{eqnarray}
\varphi=0,     \nonumber
\end{eqnarray}   
and at $z=\pm h/2$
\begin{eqnarray}
(\frac{3}{a_{33}}-\varepsilon_{0})\varphi^{'}_{z}+\frac{f_{1133}+f_{1313}}{a_{33}}\psi^{'}_{x}-\frac{f_{1313}}{a_{33}}w^{''}=0.
\nonumber
\end{eqnarray}
Here  $k=5/6$ is the shear correction factor and
\begin{eqnarray}
Q_{x}^{el}=\int_{A}E_{x}(x,z)dA.		\nonumber
\end{eqnarray}
\begin{eqnarray}
M_{x}^{el}=\int_{A}zE_{x}(x,z)dA.		\nonumber
\end{eqnarray}

Now, let us introduce new quantities
\begin{eqnarray}
&& c_{1} =\sqrt{\frac
{c_{11}-\frac{d_{31}^2}{a_{33}}-\frac{(f_{1133}+f_{1313})^2}{a_{33}r_{0}^2}}
{\varrho}}, \quad  c_{s}=\sqrt{\frac{k(c_{55}-\frac{d_{15}^2}{a_{11}})}{\varrho}}, \nonumber \\
&&r_{0} =\sqrt{\frac{I}{A}}.				\nonumber
\end{eqnarray}
Then, the dispersion relation can be written as
\begin{eqnarray}
\omega^4 -\omega^2[\frac{c_{s}^2}{r_{0}^{2}}+(c_{s}^2+c_{1}^2)k^2]+c_{1}^2 c_{s}^2 k^4=0, \nonumber
\end{eqnarray}
or
\begin{eqnarray}
&& \omega=\sqrt{\frac{1}{2}((c_{1}^2 +c_{s}^2)k^2 +\frac{c_{s}^2}{r_{0}^2})\pm  
\sqrt{\frac{1}{4}((c_{1}^2 +c_{s}^2)k^2 +\frac{c_{s}^2}{r_{0}^2})^2 -c_{s}^2 c_{1}^2 k^4}}			\label{eq:b9}
\end{eqnarray}
The equation (\ref{eq:b9}) presents the frequencies for two wave modes. The lower frequency relates to flexural wave and the higher frequency relates to shear wave.

From the formulas (\ref{eq:b9}) and (\ref{eq:0002}) it is seen that  the eigenfrequencies increase if we take into account the piezoelectric and/or flexoelectric effects. The results of the comparison between the eigenfrequencies calculated within the framework of the Euler-Bernoulli and the Timoshenko beam models coincide with the conclusions related to single-walled nanotubes presented in \cite{3}.

% !!!!!!!!!!!!!!!!!!!!!!!!!!!!!!!!!!!!!!!!!!!!!!!!!!!!!!!!!!!!!!!!!!!!!
\section{Two-dimensional model}

In this model, we consider  two components of the displacement $u=u_1$ and $w=u_3$ and only three components of strain tensor acting in $x$ and $z$ direction
\begin{eqnarray}
\varepsilon_{x}=u_{x}, \quad \varepsilon_{z}=w_{z}, \quad
\gamma_{xz}=u_{z}+w_{x}								\nonumber
\end{eqnarray}
The electric field may be written as
\begin{eqnarray}
&&-E_x =a_{11}P_x +d_{15}\gamma_{xz}+f_{1111}u_{xx}+f_{1331}(u_{xx}+w_{xz})+
f_{3311}w_{31} \nonumber \\
&&-E_z =a_{33}P_z +d_{31}u_{x}+f_{1133}u_{xz}+f_{1313}(u_{31}+w_{11})+f_{3333}w_{zz}.
												\nonumber
\end{eqnarray}
Now, the polarization may be expressed as
\begin{eqnarray}
&&P_{x}=-\frac{1}{a_{11}}(\varphi^{'}_{x}-d_{15}(u_z+w_x)-f_{1111}u_{xx}-f_{1331}(u_{xx}+w_{xz})-\nonumber \\
&&f_{3311}w_{31})\nonumber \\
&&P_{z}=\frac{1}{a_{33}}(\varphi^{'}_{z}-d_{31}u_{x}-f_{1133}u_{xz}-f_{1313}(u_{31}+w_{11})-f_{3333}w_{zz}).\nonumber 
\end{eqnarray}
In this case, the equation (\ref{eq:c1}) takes the form
\begin{eqnarray}
&&H=\frac{1}{2}a_{11}P_{x}^2 +\frac{1}{2}a_{33}P_{z}^2+\frac{1}{2} c_{11}\varepsilon_{x}^2+\frac{1}{2}c_{33}\varepsilon_z^2+\frac{1}{2}c_{55}\gamma_{xz}^2+c_{13}\varepsilon_{x}\varepsilon_{y}+ \nonumber \\
&&d_{15}\gamma_{xz}P_{x}+d_{31}\varepsilon_{x}P_{z}+
P_1 (f_{1111}u_{xx}+F_{1331}(u_{zz}+w_{xz})+f_{3311}w_{xz})+ \nonumber \\
&&P_3 (f_{1133}u_{xz}+f_{1313}(u_{xz}+w_{xx}+f_{3333}w_{zz})-\nonumber \\
&&\frac{1}{2}\varepsilon_{0} (\varphi^{'}_{x})^2-\frac{1}{2}\varepsilon_{0} (\varphi^{'}_{z})^2+\varphi^{'}_{x}P_{x}+\varphi^{'}_{z}P_{z}= \nonumber \\
&&\frac{1}{2a_{11}}(\varphi^{'}_{x}-d_{15}(u_{z}+w_{x})-f_{1111}u_{xx}-f_{1331}(u_{zz}+w_{xz})-f_{3311}w_{xz})^2\nonumber \\
&&+\frac{1}{2a_{33}}(\varphi^{'}_{z}-d_{31}u_{x}-f_{1133}u_{xz}-f_{1313}(u_{zz}+w_{xx})-f_{3333}w_{zz})^2+    \nonumber \\
&&\frac{1}{2}c_{11}u_{x}^2+\frac{1}{2}c_{33}w_{z}^2+ 
\frac{1}{2}c_{55}(u_{z}+w_{x})^2+c_{13}u_{x}w_{z}+	\nonumber \\
&&\frac{1}{a_{11}}(\varphi^{'}_{x}-d_{15}(u_{z}+w_{x})-f_{1111}u_{xx}-f_{1331}(u_{zz}+w_{xz})-f_{3311}w_{xz})\cdot\nonumber \\
&&(\varphi^{'}_{x}+d_{15}(u_{z}+w_{x})+f_{1111}u_{xx}+f_{1331}(u_{zz}+w_{xz})+f_{3311}w_{xz})+   \nonumber \\ 
&&\frac{1}{2a_{33}}(\varphi^{'}_{z}-d_{31}u_{x}-f_{1133}u_{xz}-f_{1313}(u_{zz}+w_{xx})-f_{3333}w_{zz})\cdot  \nonumber \\ 
&&(\varphi^{'}_{z}+d_{31}u_{x}+f_{1133}u_{xz}+f_{1313}(u_{zz}+w_{xx})+f_{3333}w_{zz})-
 \frac{1}{2}\varepsilon_{0} (\varphi^{'}_{x})^2-\frac{1}{2}\varepsilon_{0} (\varphi^{'}_{z})^2	\nonumber
\end{eqnarray}

The kinetic energy is
\begin{eqnarray}
T_{k}=\frac{1}{2}\int_{V_b} \varrho[\dot{u^2}+\dot{w}^2]dV_b																	\nonumber
\end{eqnarray}

The work done is
\begin{eqnarray}
W_{d}=\int_{0}^{l}(pu+qw)dx,	\nonumber							
\end{eqnarray}
where $p$ is the longitudinal load and $q$ is the transverse load.
Via the Hamilton's principle, we have the equations of motion
\begin{eqnarray}
&&u_{xx}(c_{11}-\frac{d_{31}^2}{a_{33}})
+u_{zz}(c_{55}-\frac{d_{15}^2}{a_{11}})+w_{xz}(c_{13}+c_{55}-\frac{d_{15}^2}{a_{11}})-\varphi_{xz}(\frac{d_{31}}{a_{33}} +\frac{d_{15}}{a_{11}})+ \nonumber \\
&&w_{xxx}(\frac{d_{15}}{a_{11}}f_{1111}-\frac{d_{31}}{a_{33}}f_{1313})
+\frac{f_{1331}}{a_{11}}d_{15}u_{zzz}- w_{xzz}(\frac{d_{31}}{a_{33}}f_{3333}+\frac{d_{15}}{a_{11}}f_{1331}) \nonumber \\ 
&&\frac{f_{1111}}{a_{11}}[\varphi_{xxx}+f_{1111}u_{xxxx}+f_{1331}(u_{xxzz}+w_{xxzz})+f_{3311}w_{xxxz}]\nonumber \\
&&\frac{f_{1331}}{a_{11}}[\varphi_{xzz}+f_{1111}u_{xxzz}+f_{1331}(u_{zzzz}+w_{xzzz})+f_{3311}w_{xzzz}]\nonumber \\
&&\frac{f_{1133}+f_{1313}}{a_{33}}[\varphi_{xzz}+f_{1133}u_{xxzz}+f_{1313}(u_{xxzz}+w_{xxxz})+f_{3333}w_{xzzz}]-p=\varrho \ddot{u}, \nonumber \\
&&w_{xx}(c_{55}-\frac{d_{15}^2}{a_{11}})+w_{zz}c_{33}+u_{xz}(c_{13}+c_{55}-	\frac{d_{15}^2}{a_{11}})										                                              \label{eq:a3}\\
&&-\frac{d_{15}}{a_{11}}\varphi_{xx} - \frac{d_{15}}{a_{11}}f_{3311}w_{xxz}+u_{xxx}(\frac{d_{31}}{a_{33}}f_{1313}-\frac{d_{15}}{a_{11}}f_{1111})-
\nonumber \\
&&\frac{f_{1313}}{a_{33}}[\varphi_{xxz}+f_{1133}u_{xxxz}+f_{1313}(u_{xxxz}+w_{xxxx})+f_{3333}w_{xxzz}]+\nonumber \\
&&\frac{f_{1331}}{a_{11}}[\varphi_{xxz}+f_{1111}u_{xxxz}+f_{1331}(u_{xzzz}+w_{xxzz})+f_{3311}w_{xxzz}]\nonumber \\
&&\frac{f_{3333}} {a_{33}} [\varphi_{zzz}+f_{1133}u_{xzzz}+f_{1313}(u_{xzzz}+w_{xxzz})+f_{3333}w_{zzzz}] 
-q=\varrho \ddot{w},        \nonumber \\
&&\varphi_{xx}(\frac{3}{a_{11}}-\varepsilon_{0})+\varphi_{zz}(\frac{3}{a_{33}}-\varepsilon_{0})-u_{xz}(\frac{d_{15}}{a_{11}}+\frac{d_{31}}{a_{33}})-\frac{d_{15}}{a_{11}}w_{xx}- 			\nonumber \\
&&\frac{f_{1111}}{a_{11}}u_{xxx}-\frac{f_{3333}}{a_{33}}w_{zzz}-u_{xzz}(\frac{f_{1331}}{a_{11}}+\frac{f_{1313}}{a_{33}})+	
w_{xxz}(\frac{f_{1331}}{a_{11}}+\frac{f_{3311}}{a_{11}}+\frac{f_{1313}}{a_{33}})=0	\nonumber 		
\end{eqnarray}
with the boundary conditions presented for the brevity sake in the form of partial derivatives of the electric enthalpy function as:
at $x=0$
\begin{eqnarray}
u=u_{x}^{'}=w=w_{x}^{'}=\varphi=0 ,   \label{eq:a4}
\end{eqnarray}
at $x=l$
\begin{eqnarray}
&&\frac{\partial H}{\partial u_{x}}-\frac{\partial}{\partial x}(\frac{\partial H}{\partial u_{xx}})-\frac{1}{2}\frac{\partial}{\partial z}\frac{\partial H}{\partial u_{xz}}=P_{*},\nonumber \\
&&\frac{\partial H}{\partial u_{xx}}=0, \nonumber \\
&&\frac{\partial H}{\partial w_{x}}-\frac{\partial}{\partial x}(\frac{\partial H}{\partial w_{xx}})-\frac{1}{2}\frac{\partial}{\partial z}\frac{\partial H}{\partial w_{xz}}=0,
\label{eq:a5}\\
&&\frac{\partial H}{\partial w_{xx}}=0, \nonumber \\
&&\frac{\partial H}{\partial \varphi_{x}}=0, \nonumber 
\end{eqnarray}
and at $z=\pm h/2$
\begin{eqnarray}
&&\frac{\partial H}{\partial u_{z}}-\frac{\partial}{\partial z}(\frac{\partial H}{\partial u_{zz}})-\frac{1}{2}\frac{\partial}{\partial x}\frac{\partial H}{\partial u_{xz}}=0,\nonumber \\
&&\frac{\partial H}{\partial u_{zz}}=0, \nonumber \\
&&\frac{\partial H}{\partial w_{z}}-\frac{\partial}{\partial z}(\frac{\partial H}{\partial w_{zz}})-\frac{1}{2}\frac{\partial}{\partial x}\frac{\partial H}{\partial w_{xz}}=0,
\label{eq:a6}\\
&&\frac{\partial H}{\partial w_{zz}}=0, \nonumber \\
&&\frac{\partial H}{\partial \varphi_{z}}=0, \nonumber 
\end{eqnarray}
If the flexoelectric coefficients are equal to zero, the equations (\ref{eq:a3}) coincide with the equations for plates presented in \cite{a5}.

Numerical solution to the system (\ref{eq:a3}) with the boundary conditions (\ref{eq:a4}) - (\ref{eq:a6}) allow us to find the deflection of the nano-switch, the strains, the stresses, and the electric potential accumulated in the cantilever.

\section{Conclusions}
In the present work, the nonlocal properties of the dielectric materials at nanoscale have been taken into account through the flexoelectric effect. Based on the Hamilton's principle and the electric enthalpy density with additional term describing the coupling between the polarization and the strain gradient, the equations of motion for Euler-Bernoulli and Timoshenko beam models as well as 2-D plate model have been derived. These equations,  may be used to analyse the static and dynamic behavior of cantilever nano-switch.  The formula connected the flexoelectric coefficient  and the maximum deflection of the cantilever has been  presented.

Two dimensional model allows us to analyse the electric potential accumulated in the nano-switch due to the flexoelectric effect.

%%%%%%%%%%%%%%%%%%%%%%%%%%%%%%%%%%%%%%%%%%%%%%%%%%%%%%%%%%%%%%%%%%%%%%

\end{document}